\documentclass[12pt,preprint]{aastex}

\newcommand{\myemail}{adamwr@calumet.purdue.edu}
\slugcomment{AJ, accepted for publication 15 Dec 2005}
\shorttitle{Synoptic, Multiwavelength Analysis of QSOs}
\shortauthors{Rengstorf, A.W. et al.}

\title{A Synoptic, Multiwavelength Analysis of a Large Quasar Sample}

\author{A.W. Rengstorf\altaffilmark{1}, R.J. Brunner, and B.C. Wilhite}
\affil{Department of Astronomy, University of Illinois, Urbana, IL 61801 \\
and \\
National Center for Supercomputing Applications, Champaign, IL 61820}

\altaffiltext{1}{current address: Department of Chemistry \& Physics, Purdue University Calumet, Hammond, IN 46383; \myemail}

\begin{abstract}

We present variability and multi-wavelength photometric information for the 933 known quasars in the QUEST Variability Survey. These quasars are grouped into variable and non-variable populations based on measured variability confidence levels. In a time-limited synoptic survey, we detect an anti-correlation between redshift and the likelihood of variability. Our comparison of variability likelihood to radio, IR, and X-ray data is consistent with earlier quasar studies. Using already-known quasars as a template, we introduce a light curve morphology algorithm that provides an efficient method for discriminating variable quasars from periodic variable objects in the absence of spectroscopic information. The establishment of statistically robust trends and efficient, non-spectroscopic selection algorithms will aid in quasar identification and categorization in upcoming massive synoptic surveys. Finally, we report on three interesting variable quasars, including variability confirmation of the BL Lac candidate PKS 1222+037.
\end{abstract}


\begin{document}

\section{Introduction}

A considerable amount of analysis has been carried out on various aspects of quasar variability \citep[see][for a review of past surveys]{Helfand01}. For example, \citet{Cristiani96} merged variability results from quasars in three separate photographic plate fields: SA 57 \citep{Trevese94}, SA 94 \citep{Cristiani96}, and the South Galactic Pole \citep{Hook94} to study ensemble and individual quasar variability for several hundred quasars. Recent work \citep{Vanden04,Devries05} has utilized the large number of quasars discovered by the Sloan Digital Sky Survey (SDSS) to make statements on quasar variability on an ensemble basis for $\sim\!\!10^4$ quasars with two or three data points per quasar light curve. In summary, most previous variability studies focus either on a large number of epochs for a relatively small sample of quasars, or a small number of epochs for a large sample of quasars. 

In this paper, we take a somewhat different approach. The QUEST Variability Survey \citep[QVS:][hereafter, R04b]{Rengstorf04b} contains light curves in up to four bandpasses for nearly 200,000 objects. The QVS contains 69 scans of a thin strip of high Galactic latitude equatorial sky ($-1\fdg2 \leq \delta \leq 0\fdg2$; $10^h \leq \alpha \leq 15^h30^m$) collected between 1999 February and 2001 April. Four broadband filters (camera filter order: $RBRV$) were used throughout the variability scans. Typical seeing at the site was about $2\farcs8$ and a limiting magnitude of $r = 19.06$ was reached. The filter set and observing cadence were chosen to optimize between multiple variability-driven projects, including a SNe 1a search and a recently published RR Lyrae catalog \citep{Vivas04}. The synoptic study of $\sim\!\!10^5$ light curves with several observations per lunation for several lunations per year for several years will serve as a testbed for algorithms and selection techniques for upcoming massive synoptic surveys (e.g., LSST, JDEM, Pan-STARRS). We see a need to develop robust techniques that bridge the gap between working with well-sampled individual quasars (e.g., 3C 273) and working with $10^5 - 10^6$ objects. 

We combine the synoptic data from the QVS with published spectroscopic quasar catalogs to study the variability of confirmed quasars. We use the Global Confidence Level (GCL) parameter (see R04b for complete details) to construct populations of variable and non-variable quasars. This technique identifies variable objects based on the statistical likelihood that the object is reliably varying above the photometric noise of the entire ensemble. An object's GCL value is the weighted average of confidence levels for every bandpass in which we have data for that object. In using the likelihood of variability rather than the amplitude of variability to identify variable objects, we account for increased photometric noise near the magnitude limit of the QVS. Using a critical GCL value for the determination of variability allows easy fine-tuning for completeness and efficiency based on the particular needs of the study. 

Extending our analysis to other wavelengths, we match the QVS synoptic data of the known quasars to several large-area, non-optical catalogs. First, we identify differences in global variability statistics based on quasar rest-frame time baseline and known quasar detection/non-detection in non-optical (Radio, IR, and X-ray) catalogs. Second, we look at the redshift and multi-wavelength luminosities of variable and non-variable populations of quasars. 

Finally, we develop a light curve morphology test that cleanly delineates highly variable objects into periodic and aperiodic sources. We apply this novel technique to the data from the QVS, using spectroscopically confirmed variable objects from the SDSS Third Public Data Release \citep[DR3;][]{Abazajian05} as our training sample. Such a morphology test can improve quasar-star separation among variable objects, aiding, for example, in identifying false positives in serendipitous quasar lens searches \citep[see, e.g.,][]{Pindor05}.

Throughout this paper, we assume the  standard WMAP cosmology \citep{Spergel03},  $\Omega = 1$, $\Omega_{\Lambda} = 0.73$, and H$_o = 71$  km s$^{-1}$ Mpc$^{-1}$.

\section{Known Quasars}

Between the SDSS DR3, the 2dF Quasar Redshift Survey \citep[2QZ;][]{Croom04} and the Veron-Cetty \& Veron catalog \citep[VC03;][]{VC03}, we have matched nearly $1,000$ known quasars to our light curve catalog: 751 quasars from the SDSS DR3, 614 from the 2QZ, and 130 from the VC03 catalog. Considering the overlap among these three catalogs, we have light curves for 933 unique, previously identified quasars. With nearly 1,000 quasars and roughly 25 data points per R-band light curve, our total data set (number of objects $\times$ number of light curve points) is of the same order of magnitude as studies carried out on SDSS quasars: $\sim\!25,000$ data points in this work compared to $\sim\!50,000$ data points used by \citet{Vanden04} and $\sim\!115,000$ data points used by \citet[][Table 1]{Devries05}. With somewhat fewer quasars but a larger number of light curve data points, we are investigating a different realm of phase space than earlier work. We see this as a vital step between earlier work, which often relied on highly sampled light curves for a small number of quasars (e.g., 3C273) and the upcoming spate of large, all-sky surveys which will have of order several observations per lunation for of order several years for $10^5 - 10^6$ quasars. The existing methods of light curve analyses, developed during the era of single- or few-quasar observations need to be updated to function robustly and automatically on future large data sets. 

\subsection{Matching to Known Quasar Catalogs}

The QVS contains 198,213 objects matched with SDSS DR3 photometry, 751 of which have been spectroscopically confirmed by SDSS as quasars.  We follow the SDSS convention\footnote[2]{as mentioned in the redshift status caveat on the SDSS spectral data products webpage: http://www.sdss.org/dr3/products/spectra/.} of considering only objects with a redshift confidence greater than 0.35 and which have been identified as quasars or high-redshift quasars \tt(zConf $> 0.35$ \&\& (specClass $= 3\  ||$ specClass $= 4$))\rm. As with our previous astrometric matching (R04b), objects within $2\farcs0$ were accepted as valid matches; the average astrometric offset for the matched quasars is $0\farcs43 \pm 0\farcs15$. 

\begin{figure}
\plotone{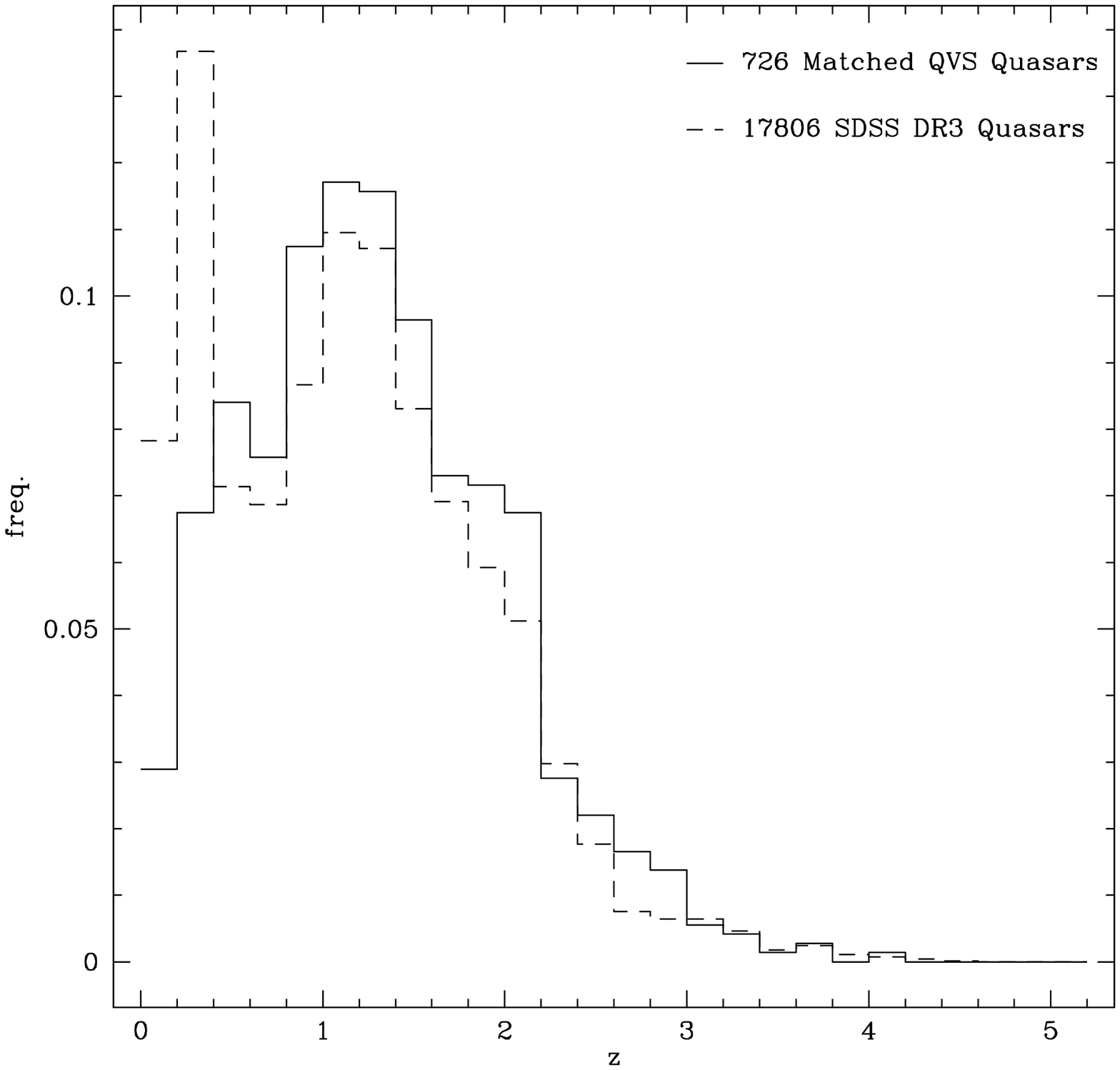}
\caption[]{The solid line shows the normalized redshift distribution of known quasars with $r < 19.06$ matched to the QVS. For comparison, the dashed line shows the normalized redshift distribution of 17,806 spectrally confirmed, $r < 19.06$ SDSS quasars.}
\label{zdistrib}
\end{figure}

Matching the entire QVS catalog to the SDSS DR3 shows that our catalog is $90\%$ complete to an SDSS $r$ magnitude of 19.06. 726 of the known quasars have $r < 19.06$. The normalized redshift distribution of these 726 quasars is given in Figure\ \ref{zdistrib} by the solid-line histogram. Figure\ \ref{zdistrib} also shows the normalized redshift distribution for 17,806 SDSS DR3 confirmed quasars brighter than $r = 19.06$ (dashed line). Note the relative paucity of low-redshift, matched quasars compared to the full SDSS sample. As previously discussed (R04b), the QUEST quasar selection techniques reject objects which are not point sources. Quasars which may have had resolved host galaxies in the QUEST scans are not expected to be present in our data. 

There are 614 high confidence quasars listed in the 2QZ catalog that were successfully matched to the QVS. Only objects with the highest 2QZ quality rating (1,1) were considered for this study.  Again, only objects with an astrometric match better than $2\farcs0$ were accepted as valid matches. The average astrometric offset for the matched 2QZ quasars is $0\farcs52 \pm 0\farcs22$.

We matched 130 quasars in the VC03 catalog, 117 of which are also seen in either the SDSS DR3 or 2QZ catalog. As with both the DR3 and 2QZ catalogs, only objects matched to better than $2\farcs0$ were considered. The average astrometric offset for the matched VC03 quasars is $0\farcs95 \pm 0\farcs49$. The increase in the VC03 mean astrometric offset and error over both the SDSS and 2QZ matching can be explained by the heterogeneous nature of the VC03 catalog. The VC03 quasars were assembled from a multitude of independent surveys, each with their own systematic and astrometric errors and biases, which result in a noticeably larger discrepancy in object astrometry compared to the more recent, homogeneous SDSS and 2dF surveys.

\subsection{Cross-matching to Other Catalogs}

The entire QVS was also matched to several publicly available, multi-wavelength surveys: the ROSAT All-Sky Survey catalog \citep[RASS;][]{Voges99,Voges00}, the VLA FIRST Survey catalog \citep{White97}, the  2 Micron All Sky Survey All Sky Data Release Point Source Catalog\footnote[3]{This publication makes use of data products from the Two Micron All Sky Survey, which is a joint project of the University of Massachusetts and the Infrared Processing and Analysis Center/California Institute of Technology, funded by the National Aeronautics and Space Administration and the National Science Foundation.} (2MASS PSC), and an SDSS type II quasar catalog \citep{Zakamska04}. Unless specifically noted otherwise, a $2\farcs0$ astrometric tolerance was used during the catalog matching.

A total of 118 objects were matched with the VLA FIRST Survey catalog with an average astrometric offset of $0\farcs70 \pm 0\farcs43$. Of these matches, 80 are among the 933 known quasars. A total of 148,204 objects were matched to 2MASS, with an average astrometric offset of $0\farcs54 \pm 0\farcs30$. Out of the 933 known quasars, 205 have 2MASS detections.

With the relatively large positional uncertainty in the RASS, it is often the case that more than one object in an optical survey lies within the error radius of an RASS object. We modified our cross-identification procedures to account for the RASS positional uncertainty of each object. Unique matches  were found within search radii of 1, 2, and 3 times an object's RASS positional uncertainty.  There were 22 uniquely matched quasars within the $1\sigma$ search radius. A total of 38 known quasars have a unique RASS match within 3 times the RASS positional uncertainty. The mean positional error for these 38 matches is $14\farcs7 \pm 7\farcs8$ with a median value of $11\farcs2$. All 38 matches were better than $33\farcs8$.  Given the sky density of the 933 matched quasars (6.94 objects per sq. deg.) and given the RASS positional uncertainty (mean positional uncertainty = $18\farcs9$), there is only a 0.0006 probability of one of the 933 known quasars falling within the mean RASS uncertainty radius by chance.

\subsection{The Data}

Variability, multi-wavelength luminosity data, and cross-matched identifiers for all 933 quasars are given in Table\ \ref{data}. The entire table is available in the electronic version of the Astronomical Journal. A subset is presented here for guidance regarding its form and content. Data presented consists of the QUEST identifier (column 1), calculated GCL and $Q_i$ values (columns 2,3), published redshifts (column 4), calculated 2500 \AA, 2 $keV$, and 2 $\micron$ luminosity densities (columns 5-7), FIRST integrated flux densities (column 8), and the cross-identifications (column 9). If a quasar was seen in more than one catalog, cross-identification is given from SDSS DR3, 2QZ, or VC03 in that order of preference. Redshift values are also taken in that order. GCL is described in the following section; $Q_i$ is explained in \S 5; the non-optical luminosity and flux densities are discussed in \S 4. The QUEST identifier can be used to obtain full light-curve information for every object in the QVS (R04b).

\section{Variability Properties of Known Quasars}

Variability is characterized herein by the Global Confidence Level (GCL), a weighted average of the variability confidence level in each of the 4 QUEST filters through which the object was detected.

\begin{equation}
GCL = \frac{\sum_{i=1}^4(N_o/N_t)_i CL_i} { \sum_{i=1}^4(N_o/N_t)_i},
\label{GCL}
\end{equation}
where $i$ is the index over the four different filters, $N_t$ is the total number of valid scans in the ensemble for a given filter, $N_o$ is the number of times the object actually appears in the ensemble for a given filter, and $CL_i$ is the variability confidence level for a given filter. The individual CL values are calculated using the $\chi^2$ probability function according to \citet{Press92},

\begin{equation}
CL\ \tbond\ P(\nu/2,\chi^2/2)\ \dbond\ [\Gamma(\nu/2)]^{-1}\!\int_0^{\chi^2/2}e^{-t}t^{(\nu/2)-1}dt,
\label{CL}
\end{equation}
where $P$ is the incomplete gamma function, $\Gamma$ is the gamma function, $\nu = N-1$ is the number of degrees of freedom, where $N$ is the number of appearances of a given object in the ensemble and $\chi^2$ is the chi-square value for the object. GCL is a measure of our confidence in the existence of variability, not the magnitude of variability. It measures only \emph{whether} a quasar is varying at a significant level above photometric noise, not \emph{how much} above the noise it is varying nor whether the quasar exhibits a bursting behavior or a smooth, monotonic change. In this work, we adopt the GCL parameter to define variable and non-variable quasar populations. A complete discussion of the GCL parameter and the variability selection criteria for the QVS is given in R04b.

\begin{figure}
\plotone{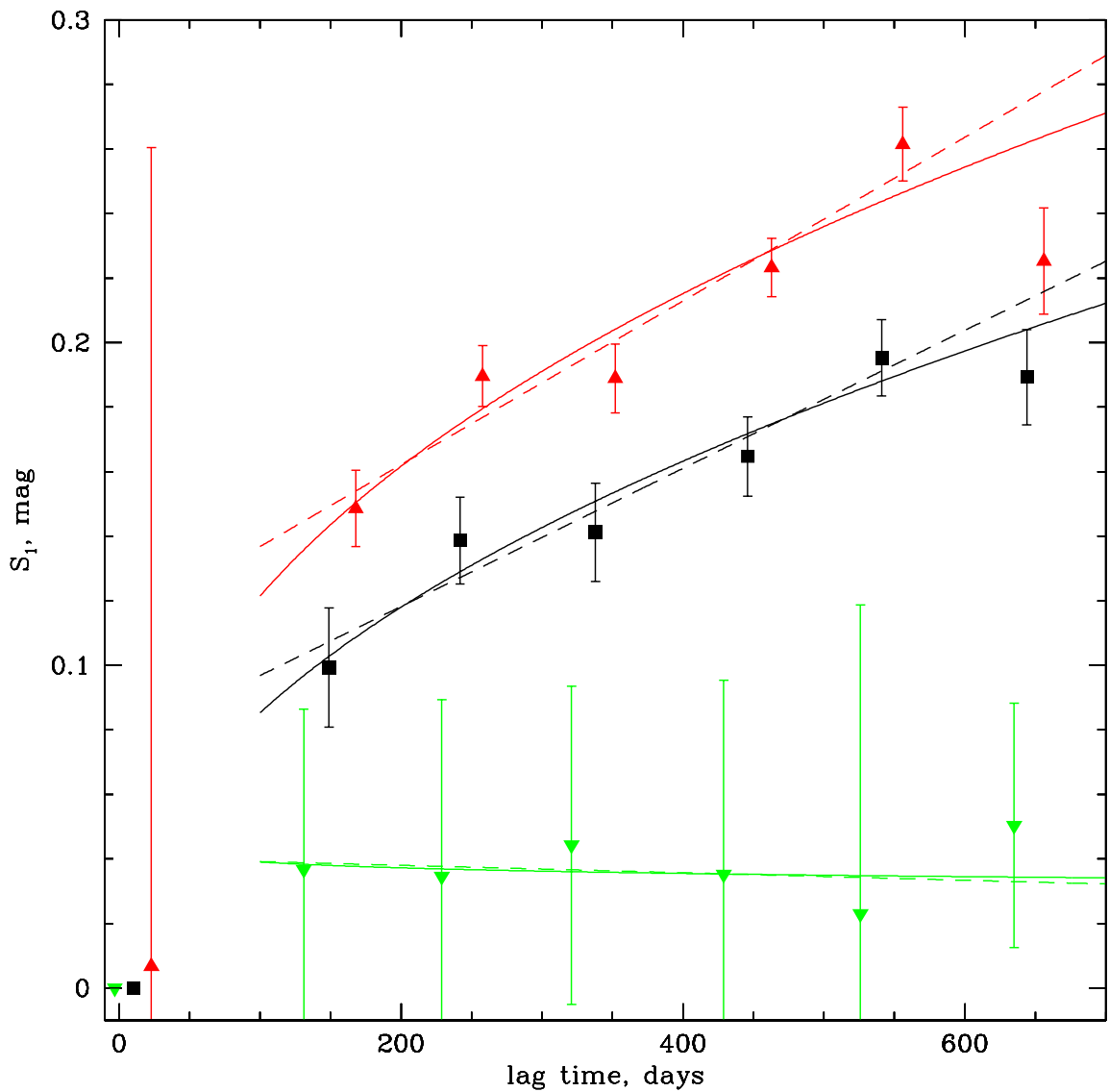}
\caption[]{First-order structure function, $S_1$, for known quasars in the QVS. Individual points are plotted at the average time lag value within each bin. Black squares show $S_1$ for all 933 quasars. Red triangles show $S_1$ for the variable (GCL $> 93$) population and inverted green triangles show $S_1$ for the non-variable (GCL $\leq 66.5$) population. Solid curves are the best power-law fit and dotted lines are the best linear fit to the data. For clarity, bin centers are shifted slightly to the right for the variable population and slightly to the left for the non-variable population.}
\label{SFplot}
\end{figure}

For comparison with other quasar variability studies, a reduced, first-order structure function is calculated for the entire population of 933 quasars. Following \citet{diClemente96}, we define the structure function as

\begin{equation}
S_1(\Delta\tau) = \sqrt{\frac{\pi}{2}\langle\vert\Delta m(\Delta\tau)\vert\rangle^2 - \langle\sigma_n^2\rangle}, \label{SF}
\end{equation}
where $\vert\Delta m(\Delta\tau)\vert = \vert m_i(t) - m_j(t+\Delta\tau) \vert$, $\sigma_n^2 = \sigma_{m_i}^2 + \sigma_{m_j}^2$, and $m_i(t)$ and $m_j(t+\Delta\tau)$ are any two data points in the quasars' light curve separated by a time $\Delta\tau$ in the quasar's rest frame. The individual error for each light curve point, $\sigma_m$, is the quadrature sum of the instrumental magnitude error and the error associated with determining the transparency of an individual exposure in the QVS (R04b, \S5.3). The factor of $\frac{\pi}{2}$ assumes the noise and photometric variability in the sample both have a Gaussian distribution. The values of $\Delta_{m}$ and $\sigma_{n}^{2}$ are binned by rest-frame time lag ($\Delta{\tau}$), in bins $100d$ in width.  The brackets indicate that those values are are averaged over each bin.  The structure function for the full sample is shown in Figure\ \ref{SFplot} (black squares).  The error bars are obtained through standard error propagation of eq.\,(\ref{SF}).  The uncertainties in the quantities $<\Delta_{m}>$ and $<\sigma_{n}^{2}>$ are simply the statistical error in the mean for the values in that bin.  We note that these errors are calculated with no attempt to account for covariance between points, though an individual quasar may contribute 50---100 values of $\Delta_{m}$ and $\sigma_{n}^{2}$ to any given bin.  A full treatment of the covariant errors is beyond the scope of this paper.  We simply present these structure functions, their errors, and the resultant fits to the data, for surface comparisons with previous work.


Figure\ \ref{SFplot} also shows the best linear (black dashed line) and power law (black solid line) fits to $S_1$. Following the idealized model for the structure function described by \citet{Hughes92}, the shortest time frame bin is consistent with the fluctuation noise floor of the structure function and therefore not used in the fitting process. The longest time frame bin contains time lags which approach the length of the entire QVS survey. Poor sampling at the time-limit of the survey can cause the structure function to turn over, as seen in Figure\ \ref{SFplot}, or otherwise act erratically \citep{Hughes92}. This data point as well was not used in the fitting process. The intermediate, well-sampled region of the structure function shows growth with increasing time lag. The power law fit was done via least-squares fitting in log-log space using bins with time lags greater than 50 and less than 600 days. Both the linear and power law fit yield acceptable results. The somewhat better linear fit yields a slope of $2.1 \pm 0.3 \times 10^{-4}$ mag/day with a reduced Chi-squared value of 1.46. The power law fit yields a power-law index of $\alpha = 0.47 \pm 0.07$ with a reduced Chi-squared value of 1.37. Assuming a functional form $(t/t_o)^\alpha$ for the power-law fit gives a characteristic time scale, $t_o, = 1.9 \times 10^4$ days.

Figure\ \ref{allperc} shows the cumulative percentage of the 933 quasars that have a GCL above a certain value, $GCL_o$, shown by the bold solid curve. Roughly $26\%$ (243) of the known quasars were seen as variable in the QUEST variability scans with $GCL_o = 99$. Almost $39\%$ (361) are variable with $GCL \geq 93$. After an initial rise due to strongly variable quasars, decreasing $GCL_o$ only gradually increases the percentage of known quasars reliably considered to vary. The break in the distribution at $GCL_o = 93$, shown by the solid vertical line, suggests an empirical criterion for considering a QVS object as significantly variable. However, Figure\ \ref{allperc} is somewhat misleading as it fails to take into account the relativistic time dilation. To correct for this, it is necessary to shift to the quasars' rest frames and to study the variability completeness as a function of redshift.

\subsection{Variability Detection as Function of Redshift}

The nominal time baseline between first and last observations for the QVS is 26 months. Data were taken between February 1999 and April 2001; however, considering the patchy RA coverage over the course of the scans, any given light curve may have a time baseline noticeably shorter than the full 26 months. The left-hand panel in Figure\ \ref{timeframes} shows a histogram of the observer-frame time baselines for the 933 known quasars. The right-hand panel of Figure\ \ref{timeframes} shows a histogram of the rest-frame time baselines, deredshifted using published redshifts. The mean quasar rest-frame time baseline for the 933 known quasars is $335 \pm 124$ days. While deredshifting works to reduce the amount of time which we are investigating, it also results in better light curve coverage as a function of time in the quasars' rest frames.

\begin{figure}
\plotone{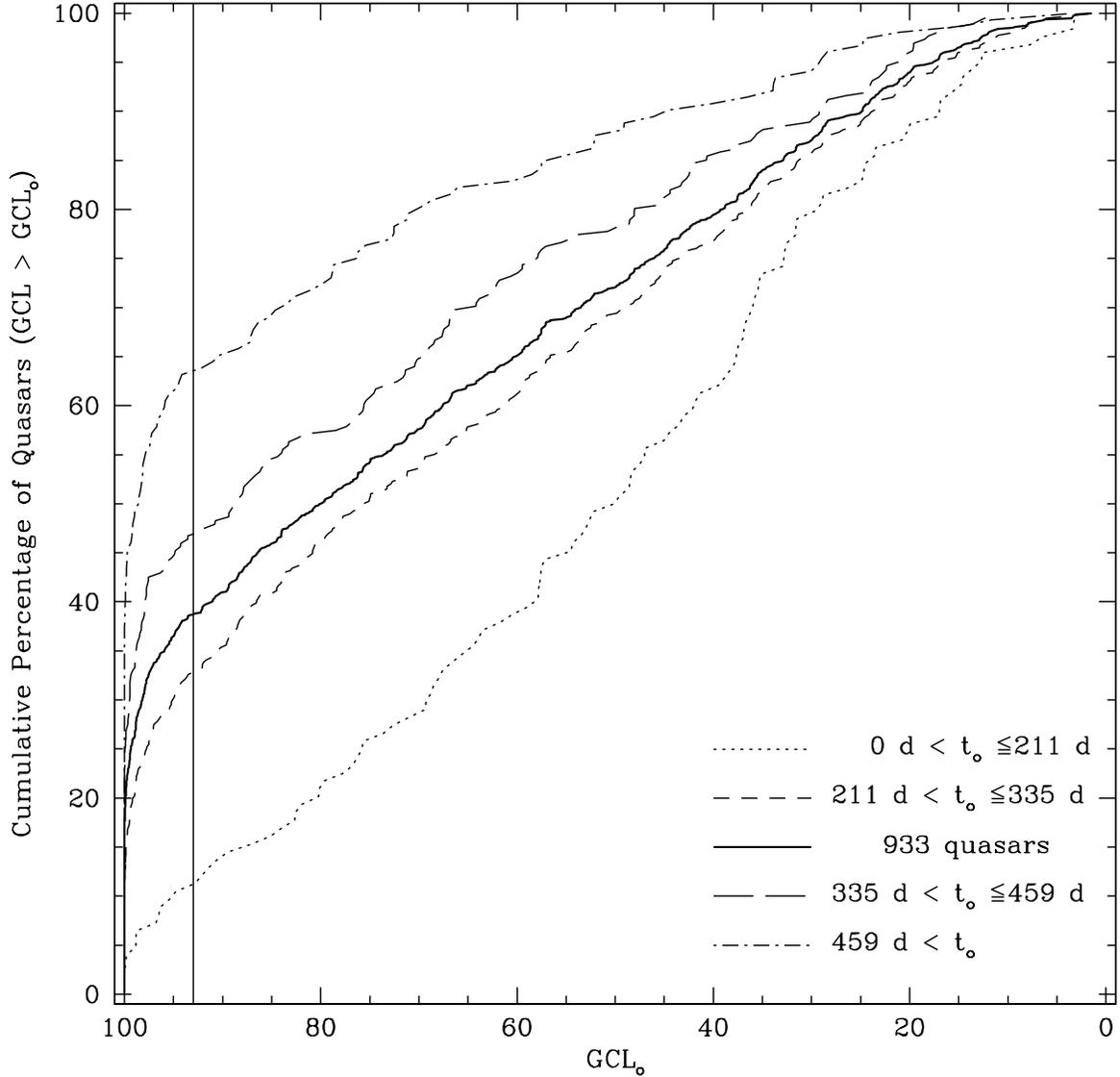}
\caption[]{Cumulative percentage of 933 known quasars which have a GCL value greater than a critical GCL value, $GCL_o$. Solid bold curve shows distribution for all 933 quasars. The dotted line shows distribution for 124 quasars with a rest frame baseline less than 211$d$. The short-dashed line shows that for 396 quasars with rest frame baselines between 211$d$ and 335$d$. The long-dashed line shows that for 261 quasars with rest frame baselines between 335$d$ and 459$d$. The dot-dashed line shows that for 152 quasars with rest frame baselines greater than 459$d$. The solid vertical line at $GCL_o = 93$ indicates the empirical cutoff for variability consideration.}
\label{allperc}
\end{figure}

\begin{figure}
\plotone{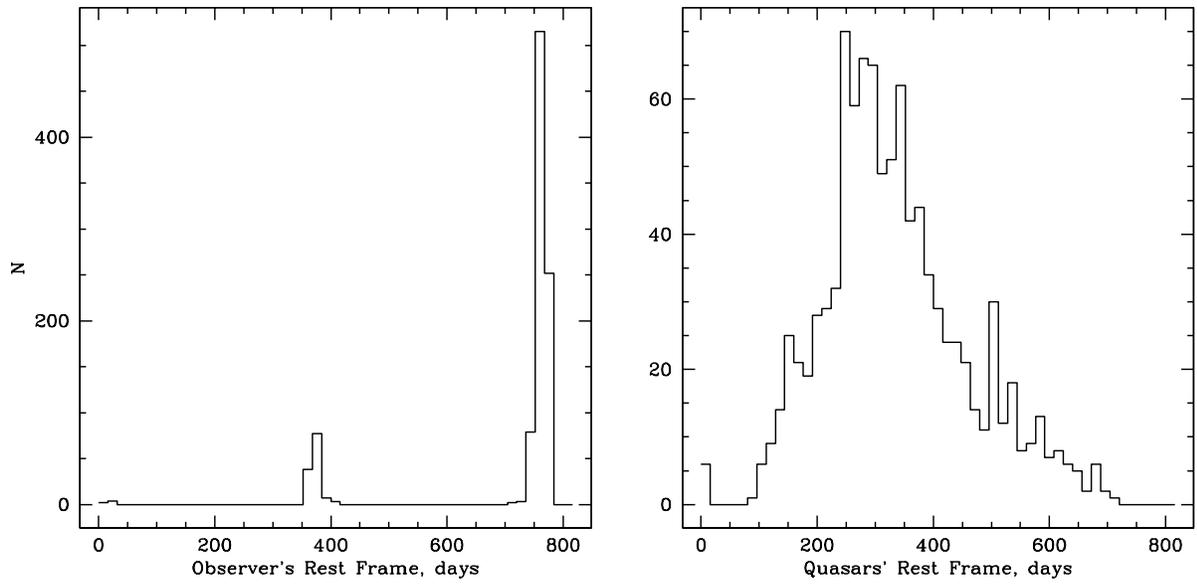}
\caption[]{Left-hand plot shows histogram of time baselines for single-bandpass light curves of 933 known quasars in the observer's rest frame. Right-hand plot shows the histogram of the same data set, corrected to the quasars' rest frames.}
\label{timeframes}
\end{figure}

\begin{figure}
\plotone{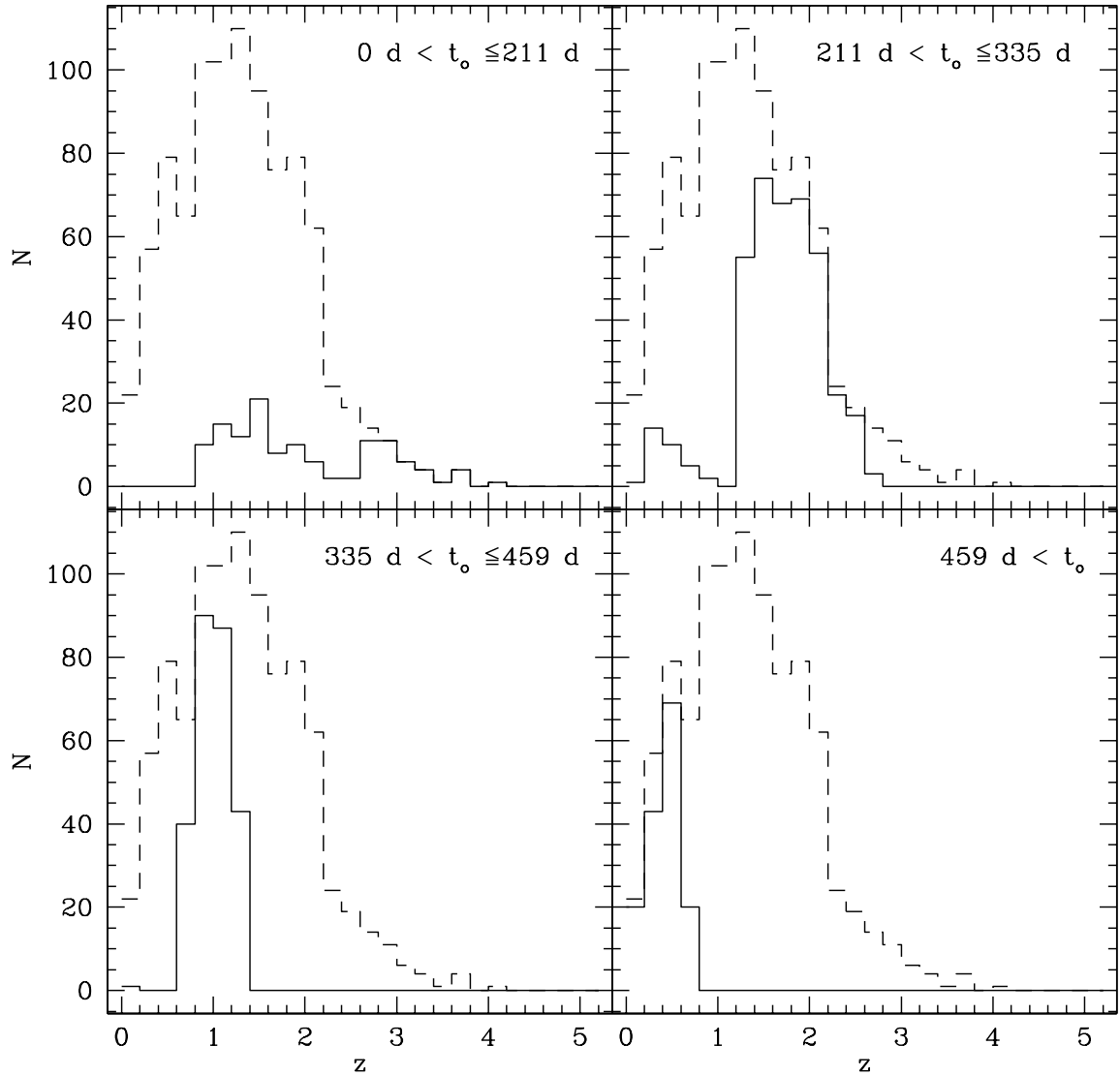}
\caption[]{Redshift distributions for each of the timeframe bins shown in Figure \ref{allperc}. Each panel contains the redshift distribution of the full population of 933 quasars shown by the dotted-line histogram.}
\label{zhistos}
\end{figure}

Figure\ \ref{allperc} also shows the cumulative distributions for quasars in various rest-frame time bins. With the range of rest-frame time baselines shown in Figure\ \ref{timeframes}, there is not an exact correspondence between time frame and redshift bins. Figure\ \ref{zhistos} shows the redshift distributions for each of the time bins in Figure\ \ref{allperc}. With the exception of the longest time baseline bin, all histograms have a low redshift tail, corresponding to the few short observer-frame time baselines shown in the left-hand plot of Figure\ \ref{timeframes}.

As expected, there is a strong correlation between rest-frame time baseline and variability detection \citep{Cristiani90,Trevese94,Cristiani96,Cristiani97,Kawaguchi99}. Over 60\% of the quasars in the longest time baseline bin have GCL $> 93$, compared to roughly 10\% of quasars in the shortest time baseline bin.  When considering the stochastic nature of quasars' light curves, there is no a priori reason for a preferred time scale for their variability. However, given the typical photometric errors in the QVS, we do see a minimum time scale for detecting variability in a significant fraction of quasars. The curves in Figure\,\ref{allperc} show that many quasars are not picked as variable with $GCL \geq 93$ until after approximately one year in the quasar rest frame. When considering quasars that were sampled for at least 15 months in their rest frame, we expect with high confidence that the majority will vary above the photometric noise.

\subsection{Definition of Variable and Non-variable Quasar Populations}

The original purpose of the GCL parameter was to identify likely variable quasar candidates within the QVS for subsequent spectroscopic confirmation \citep{Rengstorf04a}. We emphasize again that GCL is not intended as a measure of the strength of variability. It measures only \emph{whether} a quasar is varying at a significant level above photometric noise. We adopt the GCL parameter to define our variable and non-variable quasar populations. This is advantageous in that GCL takes into account the increasing photometric errors of fainter objects in the sample. This introduces a possible bias against detecting slight variability in quasars near the QVS magnitude limit, but insisting upon a high GCL for variability determination ensures a minimum of false positives in our variable quasar population. 

From investigation of Figure\ \ref{allperc} the percentage of detectably variable quasars rises sharply from $GCL_o = 100$ to a break in the cumulative distribution at $GCL_o = 93$. We see from Figure\,\ref{allperc} that this break is present in all but the shortest time frame bin and that it does not vary much  from bin to bin. This supports the choice of $GCL_o = 93$ as a good empirical limit for variable quasars in the QVS.

The 361 known quasars with GCL $> 93.0$ are defined as the \emph{variable} population. The remaining quasars are broken into populations of \emph{marginally variable} and \emph{non-variable} quasars such that the number of non-variable quasars is equal to the number of variable quasars. Figure\ \ref{SFplot} also shows the structure functions for both the variable (red triangles) and non-variable (green inverted triangles) populations. As with the full sample, both the variable and non-variable populations were fit to linear (dotted line) and power-law (solid line) functions. The non-variable population structure function is best fit by the linear relation and we note that the non-variable structure function is flat within the errors, as expected. The variable population structure function shows somewhat increased variability compared to the full sample and is best fit with a power law with index $\alpha = 0.41 \pm 0.07$. Again, using the functional form $(t/t_o)^\alpha$, the variable quasar population has a characteristic time scale of $1.7 \times 10^4$ days. The variable population power law index is equal to the full sample power law index to within their errors. The increased level of variability is seen in the shorter characteristic time scale for the power law fit over the intermediate time lag values in the QVS data. The traditional structure function shows the effectiveness of using GCL to define populations of quasars based on their likelihood of variability. We also note that the variable structure function is very close in appearance to the overall structure function, indicating that the GCL $> 93$ population is  responsible for the majority of the variability seen in Figure\ \ref{SF}. In the rest of our analysis, we only utilize the variable ($GCL > 93$) and non-variable ($GCL \leq 66.5$) populations of known quasars.  

\section{Properties of the Variable and Non-variable Quasar Populations}

\subsection{Redshift}

As mentioned above and as illustrated in Table\ \ref{props}, those quasars with lower redshift (i.e., longer rest-frame time baseline) are more likely to appear as variable in the QVS. This is entirely expected, given a survey with a fixed time frame (R04b). The entire quasar population has a mean redshift of $1.32 \pm 0.69$ with  a median value of 1.27. The variable quasar population has a mean value of $1.12 \pm 0.63$ with median of 1.04. The non-variable population has a mean value of $1.51 \pm 0.70$ with a median of 1.47. 

Previous studies have shown quasar variability to increase with time lag for at least several years. \citep[see, e.g.,][\S4]{Cristiani96}. Our survey, with its fixed observer-frame time baseline of 26 months, shows an increased likelihood of variability with longer time lags (i.e., the smaller redshifts). Previous studies have also shown that less luminous quasars show more variability than more luminous quasars \citep[e.g.,][]{Vanden04}. Since nearby quasars tend to be less luminous than more distant quasar in a flux-limited survey, this tendency also agrees with our observation that nearby quasars are more likely to be variable. The inverse trend we see between redshift and likelihood of variability is likely entirely due to the time lag and luminosity effects.

Given our current time baselines, however, we do not see the previously published trend of increasing variability with redshift, whether due to the more variable, higher frequency photons redshifting into longer wavelength bandpasses \citep[e.g.,][and references therein]{Helfand01} or to actual evolutionary effects \citep[e.g.,][]{Hook94,Cristiani96,Vanden04}.

\subsection{2500 \AA\ Luminosity}

To characterize any differences in UV/optical luminosity between our variable and non-variable populations, we calculated a 2500 \AA\ flux density for the 933 quasars. 2500 \AA\ was chosen because it is a fairly clean region for typical quasar spectra \citep{Vanden01}. Flux densities were calculated by convolving a redshifted composite quasar spectrum \citep{Vanden01} with the SDSS filter response curves, using effective wavelengths from \citet{York00}, and SDSS DR3 PSF magnitudes, corrected for Galactic extinction and shifted to the AB magnitude system\footnote[4]{SDSS magnitudes are very close to the AB system, but the $u$ and $z$ filters have a small zero point offset. See http://www.sdss.org/dr3/algorithms/fluxcal.html for a complete description.}.

The luminosity values (in units of erg s$^{-1}$ Hz$^{-1}$) reported in Table\,\ref{props} include the median and quartile values for the entire quasar sample, for the variable population, and for the non-variable population. These data show that the median luminosity of our non-variable population is more than double that of our variable population, with overlapping 1st-to-3rd quartile ranges. Non-variable quasars have a median value of $1.63 \times 10^{31}$ while variable quasars have a median of $8.11 \times 10^{30}$. A Kolmogorov-Smirnov (KS) test shows that the two samples are rejected as coming from the same parent distribution to a level of significance of 0.001. We interpret this result as a confirmation that the difference between the median luminosity densities is statistically significant. This shows that selecting quasars based on likelihood of variability returns results consistent with trends seen in quasars ranked by amplitude of variability. 

\subsection{2$\micron$ Luminosity}
\begin{figure}
\plotone{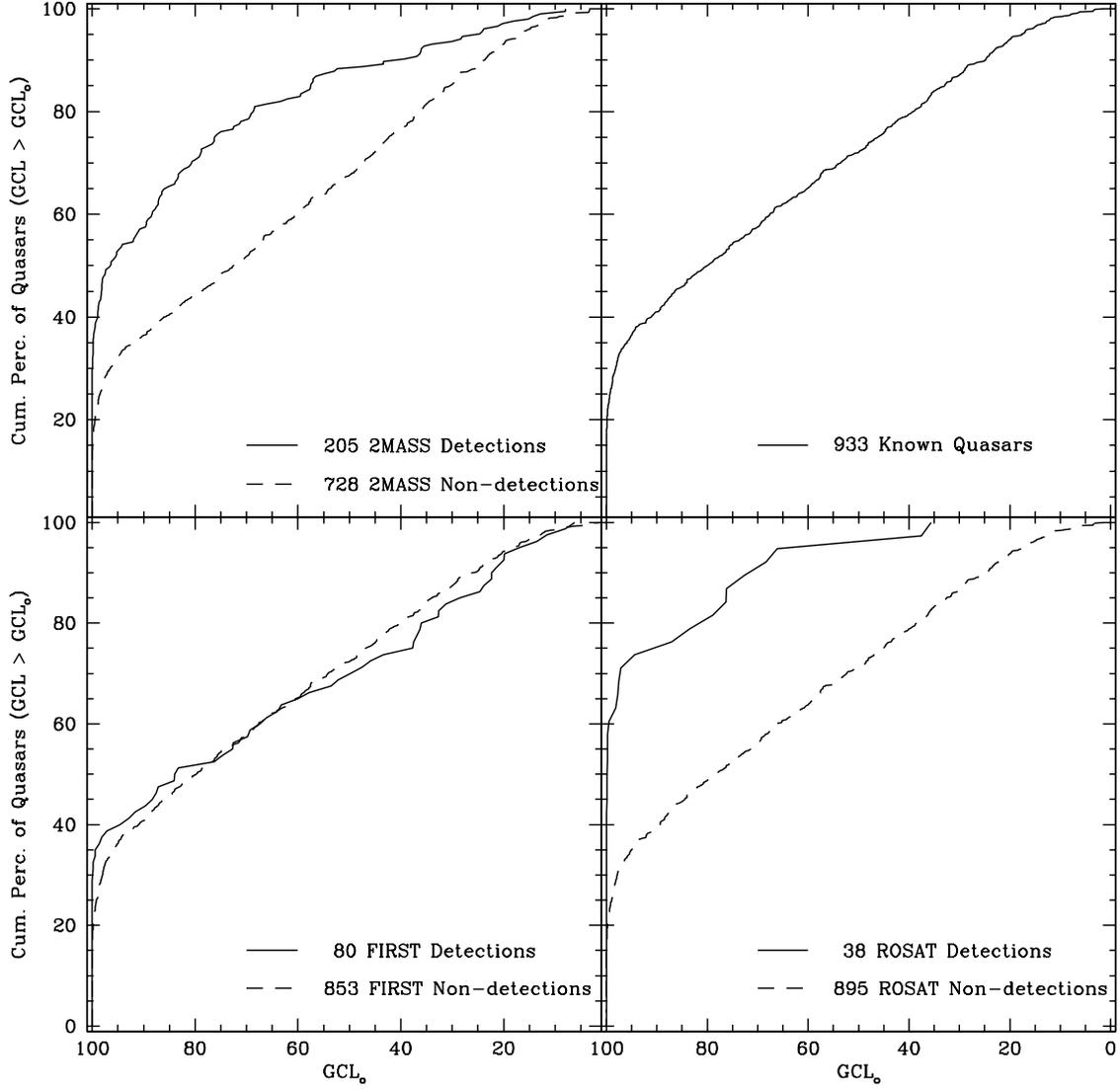}
\caption[]{ The upper-left plot shows cumulative percentage of quasars which have GCL value greater than a critical value, $GCL_o$, split into groups of 2MASS detections (solid line) and non-detections (dotted line). The lower-left plot shows the same for FIRST detections (solid line) and non-detections (dotted line) and the lower-right for unique ROSAT detections (solid line) and non-detections (dotted line). For reference, the upper-right plot shows that for the entire population of 933 quasars.}
\label{FIRSTQSO}
\end{figure}

The upper-left panel in Figure\ \ref{FIRSTQSO} shows the variability likelihood of the known quasars, split into 2MASS detections (N = 205) and non-detections (N = 728). Quasars which were detected by 2MASS have a 63\% higher chance of being seen as photometrically variable in the QVS as those not detected by 2MASS (54.5\% versus 33.4\% with GCL $> 93$). For reference, the upper-right panel in Figure\ \ref{FIRSTQSO} shows the cumulative distribution for all 933 quasars, repeated from Figure\ \ref{allperc}.

As with the 2500 \AA\ luminosity density, we calculated the luminosity density at 2$\micron$ for every 2MASS match to our data. We used an optical-infrared composite quasar spectrum from the SDSS quasars in the SWIRE ELAIS N1 field \citep{Hatz05} and the J-, H-, and K$_S$-band total response curves \citep{Cutri01}. Table\,\ref{props} reports the median, 1st quartile, and 3rd quartile 2$\micron$  luminosity density values for the full population of 205 2MASS-detected quasars, for the variable sample, and for the non-variable sample. As with the 2500 \AA\ result, the median luminosity is noticeably larger for the non-variable population than that for the variable population ($9.51 \times 10^{31}$ versus $1.43 \times 10^{31}$), again with overlapping 1st-to-3rd quartile ranges. The 2$\micron$ variable and non-variable samples' luminosity densities were also subjected to a KS test, showing that the differences between the variable and non-variable populations were significant to 99.9\%. As with 2500 \AA, the 2$\micron$ luminosity density is smaller for optically variable quasars compared to the non-variable quasars. Again, in our likelihood-selected sample of variable quasars, we see results consistent with the trend that more luminous quasars, even in the near-IR, are less optically variable.

\subsection{2 keV Luminosity}

The lower-right panel in Figure\ \ref{FIRSTQSO} shows the cumulative percentage of known quasars with GCL $> GCL_o$, split into RASS detections and non-detections. As was the case with the 2MASS detections, there is a marked difference between the two X-ray populations. Quasars which were detected by the RASS are much more likely to appear photometrically variable. This supports the suggestion that combining optical variability and X-ray flux measurements is an efficient quasar selection technique \citep[e.g.,][]{Sarajedini03,Brandt04}.

A hard X-ray ($0.5 - 2.0 keV$) flux is calculated from the published RASS counts per second using the PIMMS v3.5 software application \citep{Mukai93} with $\Gamma = 2.0$, $\alpha = -1$, and the weighted average neutral hydrogen column density calculated from HEASARC's nH application using the \citet{Dickey90} HI map. From the hard X-ray flux, the rest-frame 2 keV flux density is determined, which was used to determine the 2 keV luminosity density. Of the 38 RASS detections, 28 are in the variable population and only 3 are in the non-variable population. Table\,\ref{props} reports the median, 1st quartile, and 3rd quartile luminosity densities. Again, the non-variable population has a higher median luminosity than the variable population ($3.92 \times 10^{26}$ versus $1.69 \times 10^{26}$). This result provides a tentative suggestion for the existence of an anti-correlation between X-ray luminosity and optical variability; but with only three non-variable X-ray detected quasars, additional observations are required for a more definitive statement.

\subsection{Radio Properties}

The lower-left panel in Figure\ \ref{FIRSTQSO} shows the cumulative percentage of quasars with GCL $> GCL_o$, split into FIRST detections (N = 80) and non-detections (N = 853). The samples exhibit roughly the same cumulative GCL distribution, both tracing the overall distribution shown in the upper-right panel of Figure\ \ref{FIRSTQSO}. This finding is consistent with the comparison between FIRST detections and non-detections in the SDSS by \citet{Vanden04}.

To quantify any difference in radio flux between the variable and non-variable populations, we looked at the radio-detected quasars within each group. An equal number (N = 32) were variable as were non-variable. Table\ \ref{props} reports the FIRST flux density median, 1st quartile, and 3rd quartile values. Unlike the IR and X-ray detections, the radio loud quasars show a correlation between integrated FIRST flux density and optical variability. The non-variable population has a median integrated radio flux of 3.96 mJy, while the variable population has a median of 9.64 mJy. A KS test shows that the radio flux values for the variable and non-variable populations are reliably rejected as coming from the same parent distribution. This is in agreement with earlier works that report radio-loud quasars are more variable \citep[e.g.,][]{Vanden04,Helfand01}.

\section{Light Curve Morphology}

Using a light curve morphology test, our variable population may be separated into populations which are likely to be quasars (and other aperiodic variables) and those likely to be periodic variables. The light curves from different filters for each object are analyzed individually. The results are then averaged together, similar to the method used for calculating the GCL from the individual CLs for a given object, as detailed in R04b. The light curve morphology test is predicated on the theory that a variable quasar will tend to vary little on the time scale of an individual QUEST observing season (six to eight weeks), but may vary significantly between observing seasons. The opposite will tend to be true for shorter-term periodic variable stars, which will tend to exhibit a higher level of variability during a single observing season, with the annual average magnitude remaining fairly constant from one year to the next. This is effectively a null detection test, similar to proper motion cuts to detect quasars. By rejecting variable objects which are detectably periodic, we still keep in consideration aperiodic variable objects, which will include those stellar variables which do not show periodic behavior, those whose periodicity was missed due to insufficient time sampling, and those which have periods longer than those sampled in the QVS.

\begin{figure}
\plotone{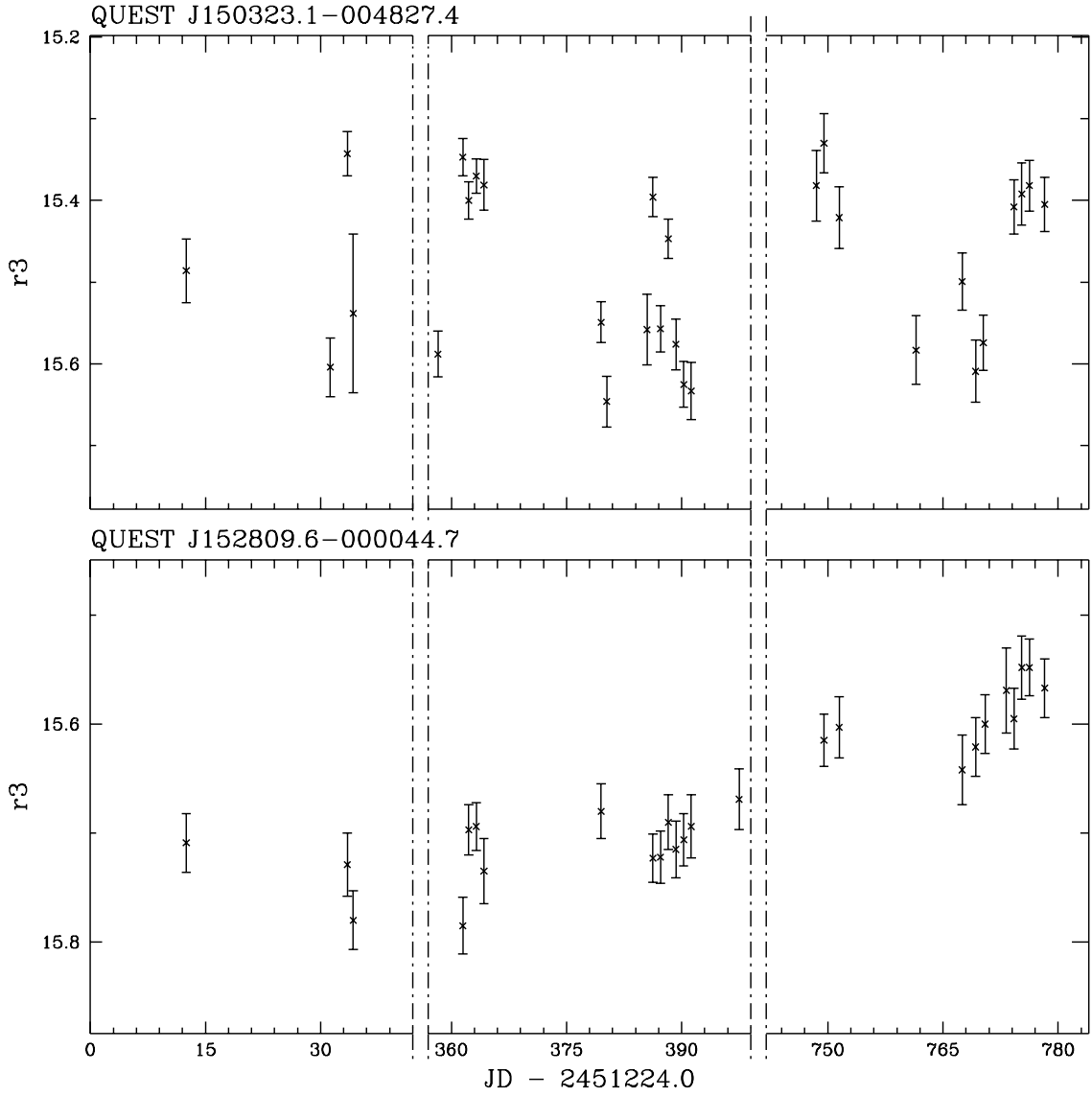}
\caption[]{Top panel shows an R-band light curve for a typical periodic variable star (GSC 05000-00417). Bottom panel shows an R-band light curve for a typical variable quasar (SDSS J152809.55-000044.8).}
\label{lcex}
\end{figure}

A light curve morphology parameter, denoted by $q_i$, is calculated. $q_i$ is the ratio of two indices, both of which exploit the qualitative difference between periodic and aperiodic variables:  a variance index, $V_i$, and a magnitude difference index, $\Delta m_i$. $V_i$ is the average of the ratio of the global variance to the single observing season variance, weighted by the percentage of total data points in a given observing season.

\begin{equation}
V_i = \frac{\sum_{i}N_i (\sigma/\sigma_i)^2} { \sum_{i}N_i}, \label{varind}
\end{equation}
where $N_i$ is the number of light curve points within a given observing season, $\sigma$ is the standard deviation of the light curve points over the entire light curve $\sigma_i$ is the standard deviation of light curve points within a given observing season, and the index $i$ runs over the three individual observing seasons. Short-term periodic variables will see larger variations within an observing season, so $\sigma_i$ and $\sigma$ will be comparable, leading to smaller values for $V_i$. Visual inspection of the representative light curves in Figure\,\ref{lcex} shows that an aperiodic light curve will tend to have a larger $V_i$ than a periodic variable.

$\Delta m_i$ compares the global standard deviation to the largest difference in mean magnitude between any two observing seasons.

\begin{equation}
\Delta m_i = \frac{\sigma} {\Delta\!<\!m\!>_{max}}, \label{magind}
\end{equation}
where $\sigma$ is again the global standard deviation and $\Delta\!<\!m\!>_{max}$ is the largest difference between single-observing-season average magnitudes. Another visual inspection of Figure\,\ref{lcex} shows that  an aperiodic light curve will tend to have a smaller $\Delta m_i$ than a periodic variable.

The $q_i$ value is formulated by dividing eq.\,(\ref{varind}) by eq.\,(\ref{magind}), giving aperiodic light curves larger values and periodic light curves smaller values:

\begin{equation}
q_i = \frac{V_i}{\Delta m_i}. \label{qindex}
\end{equation}

The $q_i$ values from each of the broadband filters are then averaged together for a global index, denoted by $Q_i$. As with the GCL calculation, the $Q_i$ is weighted by the fraction of valid scans in which the object was detected.

\begin{equation}
Q_i = \frac{\sum(N_o/N_t) q_i} { \sum(N_o/N_t)},
\label{Qindex}
\end{equation}
where the summation runs over the four different filters, $N_t$ is the total number of possible scans in the ensemble for a given filter, $N_o$ is the number of times the object actually appears in the ensemble for a given filter, and $q_i$ is the value for that filter, as calculated in eq.\,(\ref{qindex}).

\begin{figure}
\plotone{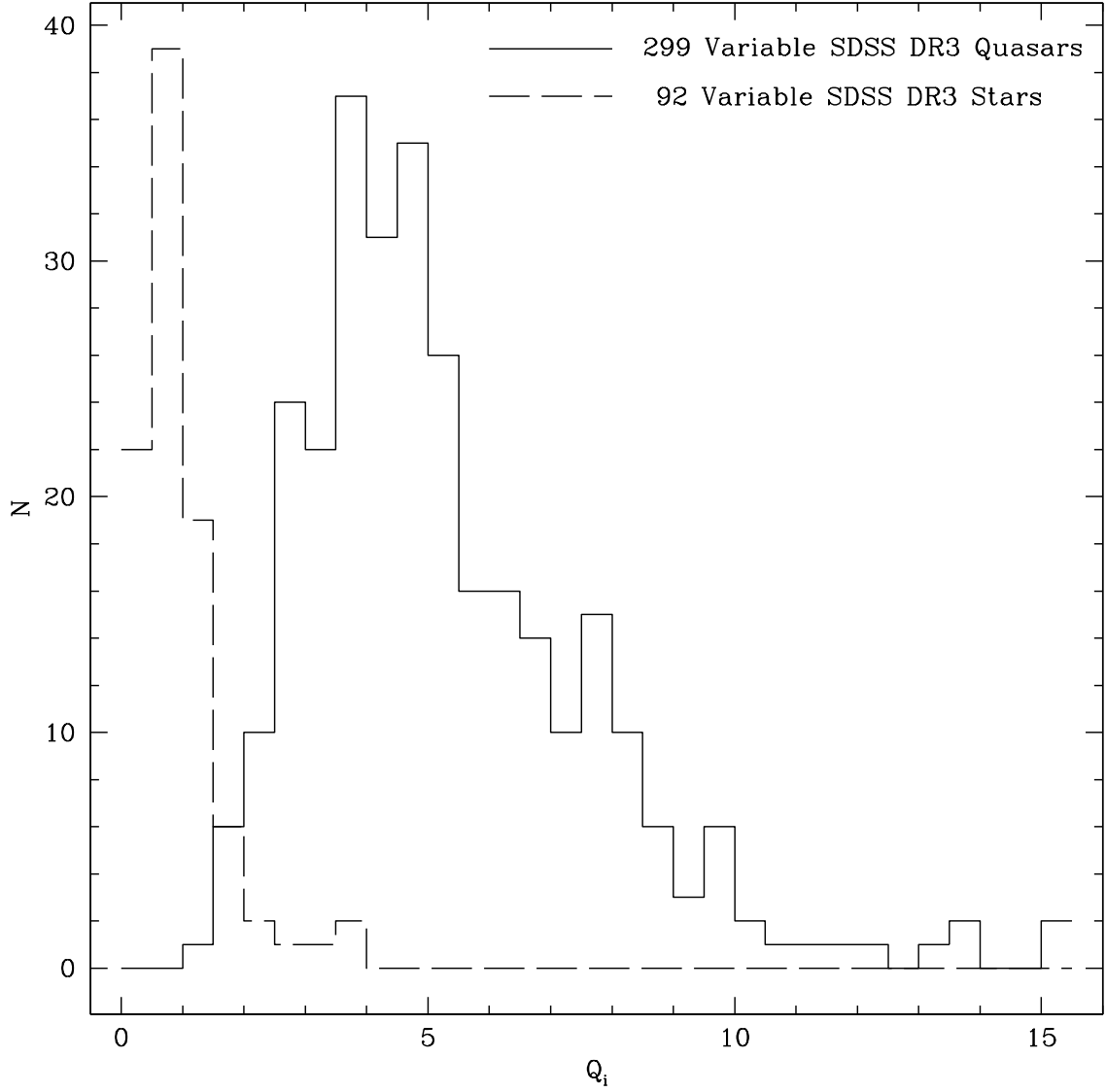}
\caption[]{Histogram showing the $Q_i$ distributions for variable (GCL $> 93$), spectrally confirmed objects from the SDSS DR3; 99 spectrally verified stars (dashed line) and 299 spectrally verified quasars (solid line).}
\label{Qhisto}
\end{figure}

As an initial test of the efficacy of our light curve morphology classifier, we calculated $Q_i$ for every variable source, which is defined here to be any object in the QVS which has a GCL $> 93$ and a corresponding SDSS spectral identification. There are 299 variable SDSS quasars \tt(specClass $= 3\  ||$ specClass $= 4$)\rm and 92 variable SDSS stars \tt(specClass $= 1$) \rm included in this test. Fig.\,\ref{Qhisto} shows a histogram of the 391 SDSS DR3 spectral objects. Only 6 out of the 92 variable stars ($6.5\%$) have $Q_i > 2.0$. Likewise, only 7 of the 299 variable quasars ($2.3\%$) have $Q_i < 2.0$.  The low-$Q_i$ variable quasars presumably fall below the cutoff due to insufficient sampling of their light curves. Of the 6 high-$Q_i$ variable stars, one (QUEST J121727.5-014036.5) is a previously unstudied variable star (SDSS J121727.4-014036.9) and requires additional observation, one (QUEST J143500.2-004605.8) is the SU Uma-type dwarf nova OU Vir \citep{Downes01} caught in outburst during the 2000 observing season, and 4 were RR Lyrae stars, cataloged during the QUEST RR Lyrae Survey \citep{Vivas04}. The RR Lyrae stars in question happened to be sampled during the 1999 observing season such that their full intra-epoch range in magnitude was not observed, causing the objects to appear to vary in mean magnitude between the 1999 and subsequent observing seasons. In short, these 6 variable stars fell above the cutoff due to either insufficient sampling of their light curves, or due to their periodicity occurring over timescales longer than the QVS.

Considering only the SDSS DR3 spectra, a cut at $Q_i = 2$ is $97.7\%$ complete and $98.0\%$ efficient at separating variable objects into groups of periodic (i.e., stars) and aperiodic (i.e., quasars) variables.   
If we expect that $\sim 10\%$ of all variable stars fall above the $Q_i = 2$ cutoff, the efficiency will diminish as the number count of stars increases with respect to the number of quasars, dropping to $\sim50\%$ efficiency when the stellar population is $\sim10$ times larger than the quasar population. It should be stressed that this is a secondary cut, meant to increase the efficiency of the primary variability cut, as detailed in R04b. We show that it can increase the efficiency of the quasar candidate list without greatly reducing the completeness.

Of the 198,213 objects in the QVS, 1,610 have GCL $> 93$. Among those objects, 624 have $Q_i > 2.0$. Of these 624 candidates, 351 are among the 933 known quasars. This indicates that the combination of GCL and $Q_i$ returns a list of quasar candidates that is \emph{at least} $56.3\%$ efficient. A minimum efficiency is reported because it is not known \emph{a priori} whether there are unconfirmed quasars among the remaining 273 objects. Recall there is a total of 361 previously known quasars with GCL $> 93$. This corresponds to a completeness of $97.2\%$ among the known variable quasars. In considering the \emph{entire} population of known quasars, using GCL and $Q_i$ cuts returns 351 out of 933 quasars, or a $37.6\%$ complete sample, with the 26-month time baseline. The completeness of the quasar candidate list, as calculated from the previously known quasars, decreased from $38.7\%$ using only the GCL cut to $37.6\%$ using GCL plus $Q_i$. However, the efficiency increases from \emph{at least} $22.4\%$, considering only the GCL of only the known quasars, to \emph{at least} $56.3\%$ considering GCL plus $Q_i$. Table\,\ref{stats} summarizes the GCL and $Q_i$ cuts and reports their effectiveness at recovering known quasars and their minimum efficiency with respect to the known quasars.

\section{Discussion}

Our current data set is trained upon known quasars which were predominantly found via traditional color-selection methods. We are therefore limited by both our variability selection biases (e.g., relatively fewer low-z quasars due to a point source requirement) and the color-selection techniques. 
Nevertheless, our sample of 933 confirmed quasars allows us to make substantive statements regarding the contrasting behavior of our variable and non-variable populations. Given the time-limited nature of the QVS, we see that 36\% of quasars reliably show optical variability with $GCL_o = 93$ after $\sim$26 observer-frame months. Correcting these data to the quasars' rest frames, we see that a majority of quasars reliably exhibit variability in our data after $\sim$15 months. We do not, however, suggest this as evidence of any quasi-periodicity or of a preferred time scale. We also see that loosening our variability criteria does little to increase the variable quasar population. There is a definite break in the cumulative distribution of variable quasars at GCL$ = 93$. This break is seen at roughly the same GCL value in all time lag bins of reasonable length and was used to define our variable  quasar population. The first-order structure functions (Figure\ \ref{SFplot}) of the variable and non-variable quasar populations show GCL to be an effective method for separating quasars for further study.

In a time-limited survey, the longest rest-frame time lags correspond to the most nearby objects, resulting in our observed anti-correlation between redshift and variability. We note that this as an inherent bias and do not make any substantive claims as to the supposedly evolutionary correlation between redshift and variability seen by others. Our smaller number of quasars do not allow the fine binning in redshift-magnitude-time lag space as used by \citet{Vanden04} to decouple these competing effects.

With a range of over 4 orders of magnitude, the distribution of 2500 \AA\ luminosity density for the entire quasar sample, as well as for the variable and non-variable sub-samples, deviates significantly from Gaussian, complicating their statistical analyses. Nevertheless, the luminosity density median values of the variable and non-variable quasars are consistent with the anti-correlation between variability and luminosity seen in earlier work; this result is supported by the KS test run between the two samples. The fact that optical/UV variability-luminosity anti-correlation is mirrored in both the IR and X-ray regimes supports the standard unification model. 

We suggest this similar behavior arises from the fact that the IR and X-ray emission are at least partially coupled to the UV/optical accretion disk emission, as described by the generally accepted model for active galactic nuclei \citep[see, e.g.,][]{Antonucci93}. IR radiation presumably originates in the outer, cooler portions of the accretion disk and in the torus from re-processed disk emission. X-ray emisson is due to UV photons from the inner accretion disk being up-scattered by relativistic electrons in the corona. Any correlations between intrinsic variations and luminosity originating in the inner accretion disk would be expected to also be seen in both the IR and X-ray regimes. A larger sample of X-ray detections and a robust strength of variability parameter, capable of reliably characterizing an individual quasars' photometric activity, are required to further investigate this multi-wavelength trend.

The radio data, however, show a qualitatively different behavior: a positive correlation between optical variability and radio flux. Blazars tend to be primarily both radio-loud and highly variable \citep[see, e.g.,][and references therein]{Antonucci93,Ulrich97}. Within the standard unification model, blazars and other core-dominant radio sources are seen when our line of sight aligns with relativistic jets perpendicular to the accretion disk. The physics and continuum emission mechanisms for radio-loud AGN originate with the relativistic jets \citep[][and references therein]{Ulrich97} and are fundamentally different from the behavior seen in other quasars. Our radio-loud quasars are not all necessarily blazars; but we suspect that there are enough blazars in our sample to give rise to the observed correlation between radio flux and optical variability.

As more quasars are observed for longer periods of time, the quantity and quality of these types of analyses will continue to increase. With robust light curves for larger numbers of quasars, we are able to explore the unification model in novel ways and can also take advantage of the unique aspects of accretion disk physics to refine variability selection techniques for quasars and other active galactic nuclei. For example, by looking at \emph{how} a quasar has varied, rather than merely by how \emph{much} it varies, we see new ways to discriminate quasars from other variable point sources. The $Q_i$ cut is very efficient at finding aperiodically variable objects (i.e., quasars) amongst periodic variables. This is an appealing avenue of study. As quasar detection efficiency increases, the amount of time required for spectroscopic follow-up decreases. We can also answer remaining questions about the unification model by finding and analyzing the variability properties of a larger sample of Type II quasars. If the inclination angle of a host galaxy or our line of sight through an obscuring torus does indeed affect the level of observed variability in a large quasar population, this would help fill in some of the remaining gaps in unification schemes.

\subsection{Interesting Objects}

There was one match between the QVS and the Type II quasars reported by \citep{Zakamska04}: QUEST J133633.7-003936.0 was matched to an SDSS DR1 Type II quasar (SDSS J133633.65-003936.4). QUEST J133633.7-003936.0 is a faint (SDSS $r$ = 19.2), non-variable (GCL = 17.85) nearby (z = 0.416) object. This quasar was observed over all three epochs of the QVS and has a rest-frame time baseline of 545 days. Its long rest-frame time baseline puts it among the quasars most likely to be seen as variable in the QVS (the upper-most curve in Figure.\,\ref{allperc}). However with its low GCL, this quasar is less likely to be variable than 98\% of the quasars in the longest time baseline bin. This quasar has somewhat red colors ($u-g = 0.92; g-r = 1.36$; $r-i = 0.41$; $i-z = 0.21$) and is missed both by standard photometric selection techniques \citep[see, e.g.,][Fig. 7]{Richards02} and by our variability technique. If the optical depth of the obscuring material is greatest towards the nuclear region of the AGN, then the region responsible for the origin of the quasar's variability would be preferentially obscured, damping the photometric variability. We would therefore expect that Type II quasars are less likely to be detected as being variable than Type I quasars. However, a sample size greater than one Type II quasar is needed to make a substantive claim to this effect.

\begin{figure}
\plotone{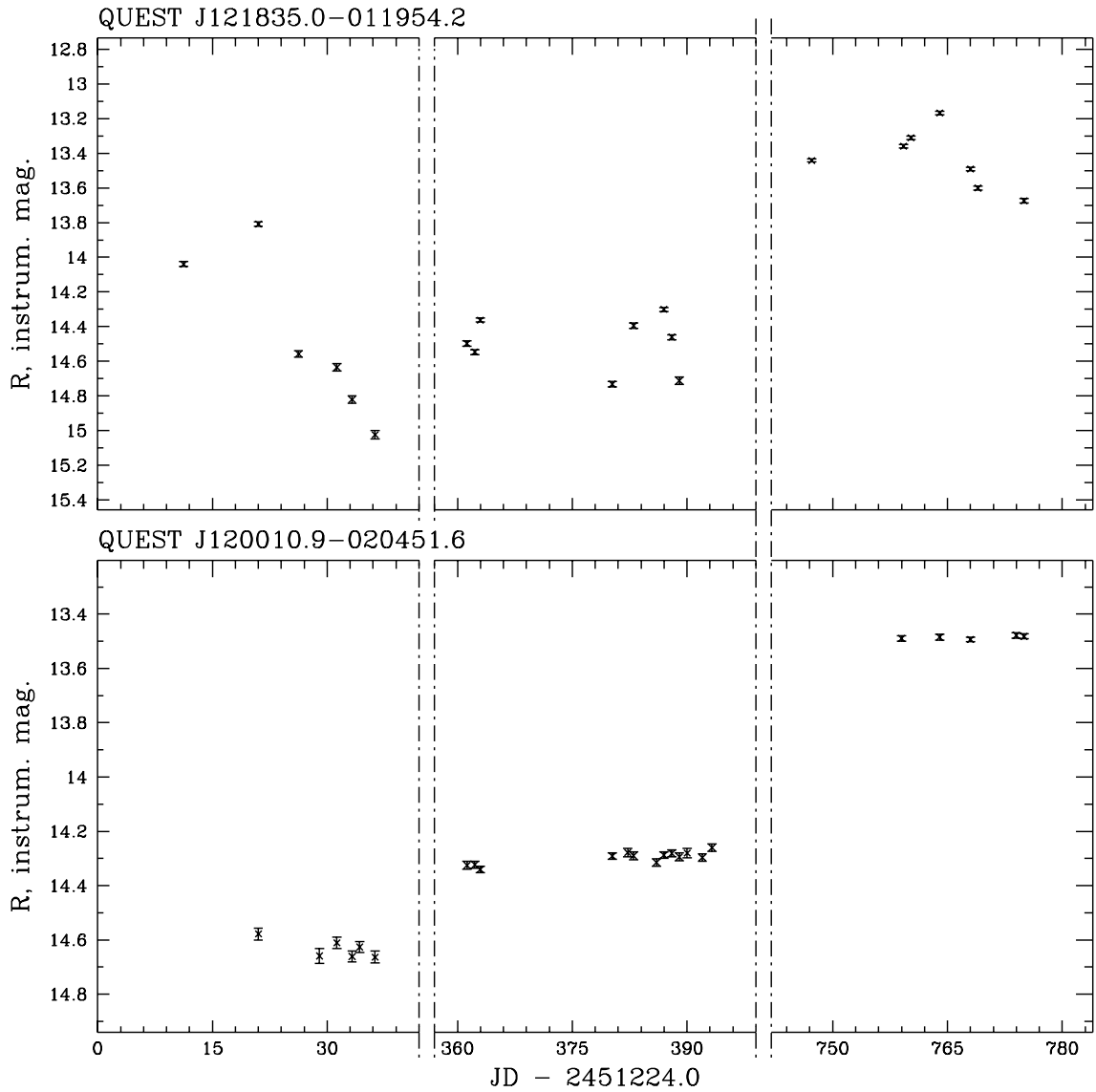}
\caption[]{QVS light curves for two highly variable, previously identified quasars: PKS 1216-010, suggested to be variable due to earlier polarimetry study and shown to vary by nearly 2 mag in the QVS and SDSS J120010.93-020451.8, which shows $\sim1$ mag of variation in the QVS.}
\label{2QSOs}
\end{figure}

Two quasars were seen as high variability outliers in the structure function analysis: QUEST J121835.0-011954.2 and QUEST J120010.9-020451.6. The quasar QUEST J121835.0-011954.2 was matched to an SDSS point source (SDSS J121834.93-011954.3), earlier identified as PKS 1216-010. It is a moderately bright (SDSS $r$ = 17.6), variable (GCL = 100; $Q_i = 6.40$), nearby (z = 0.415) object. This quasar was observed over all three epochs of QVS and shows both considerable intra- and inter-epoch variability. This quasar was part of an optical polarization study by \citet{Sluse05} and was mentioned as a candidate for optical variability due to its variable polarization level between March and May 2002. The QVS light curve certainly bears out this prediction, and we note that the QUEST data actually precedes the polarization observations by a small margin. The upper panels in Figure\ \ref{2QSOs} show an R-band instrumental-magnitude light curve for this quasar. The strong variability on both long and short time scales suggest that this object was correctly identified as a BL Lac object. 

QUEST J120010.9-020451.6 is another moderately bright (SDSS $r$ = 17.4), highly variable (GCL = 100; $Q_i = 57.05$), nearby (z = 0.09) quasar, also identified by the SDSS (SDSS J120010.93-020451.8). This object was seen as a point source in the QUEST scans, but was resolved and targeted as a galaxy by the SDSS photometry. Following SDSS spectroscopy, it was classified as a quasar. It shows a steady increase in brightness over the three QVS epochs, but little variability in any single epoch. The quasar brightened by over one a magnitude in both R and V filters in the QVS data. The lower panels in Figure\ \ref{2QSOs} show an R-band instrumental-magnitude light curve for this quasar. The lack of short time scale variability compared to the inter-epoch variability suggests this is a typical Type I quasar.

\section{Conclusions}

We have used a unique, synoptic dataset, federated with large-scale optical and non-optical public survey data, to explore quasar variability in a realm of phase space not previously examined. We have larger numbers of light curve points than recent work using SDSS data and we are spanning more area than earlier pointed observations. Using data from the QVS and known quasars from the SDSS, 2QZ, and VC03 catalogs, we have assembled 933 quasar light curves in multiple bandpasses. Using the GCL parameter from R04b, we defined populations of variable and non-variable quasars. Converting the light curves to the quasars' rest frames, we find a bias towards nearby objects being more likely to vary in a time-limited synoptic survey, likely due to luminosity and time lag effects. This trend will need to be accounted for in the upcoming synoptic surveys.

The anti-correlation between variability and redshift is evidence that we are sampling time lags in which quasar variability is still rapidly increasing with time lag. Previous work using ensemble structure function analysis has shown that quasar variability is expected to increase with time lag over at least several years \citep[e.g.,][]{Hook94,Cristiani97,Vanden04,Devries05}. Recent work has indicated a positive correlation between redshift and variability, decoupled from the correlation between variability and photon wavelength \citep{Vanden04}. We did not see the redshift-variability correlation in our data, which is not surprising given our sampled time baselines and the reported weakness of the redshift-variabilty correlation.

A near-UV (2500 \AA) luminosity density was calculated for every known quasar. The median luminosity densities for the variable and non-variable quasar populations showed an anti-correlation with likelihood of variability, consistent with earlier work \citep[e.g.,][]{Hook94,Trevese94,Cristiani96,Vanden04,Devries05}. The known quasars were also matched to the FIRST, 2MASS, and ROSAT surveys. Correlations between 2MASS and ROSAT detection and optical variability were found. Those quasars which were detected in either the IR or the X-ray were significantly more likely to be seen as variable than those quasars not detected. In each of the non-optical surveys, the detected sub-sample was then divided into variable and non-variable populations. Both the IR- and X-ray-detected populations showed that the variable population had a lower median luminosity than the non-variable population. 

Unlike the IR and X-ray surveys, the FIRST survey showed no difference in detection between the variable and non-variable populations. The median integrated radio flux density, however, was larger for the variable population than the non-variable population, also in contrast with the other non-optical surveys. \citet{Vanden04} reported the same behavior and suggest blazars as a possible explanation. 

A simple light curve morphology analysis shows that the unique, aperiodic time signature inherent to quasars can be utilized to further refine variability selection techniques. The $Q_i$ parameter efficiently separates variable point source objects in our data into groups of periodic (i.e., stellar) and aperiodic (i.e., quasars etc.) objects. While somewhat fine-tuned to the particulars of the QVS, this is an appealing test, as it is easy to implement, takes advantage of the light curve details, and is fairly robust to unevenly sampled data. These techniques used on the QVS are a useful testbed for techniques to identify and characterize quasars using synoptic photometric data without the need for follow-up spectroscopy.

\acknowledgments

The authors would like to acknowledge support from NASA through grants NAG5-12578 and NAG5-12580 as well as support through the NSF PACI Project. AWR thanks A.D. Myers and B.F. Lundgren for insightful discussions which improved the overall quality of this paper.

The authors made extensive use of the storage and computing facilities at that National Center for Supercomputing Applications and would like to thank the technical staff for their assistance in enabling this work to proceed.

The QUEST data used in this paper is based on observations obtained at the Llano del Hato National Astronomical Observatory, operated by CIDA for the Ministerio de Ciencia y Tecnologia of Venezuela.

Funding for the creation and distribution of the SDSS Archive has been provided by the Alfred P. Sloan Foundation, the Participating Institutions, the National Aeronautics and Space Administration, the National Science Foundation, the U.S. Department of Energy, the Japanese Monbukagakusho, and the Max Planck Society. The SDSS Web site is http://www.sdss.org/.

The SDSS is managed by the Astrophysical Research Consortium (ARC) for the Participating Institutions. The Participating Institutions are The University of Chicago, Fermilab, the Institute for Advanced Study, the Japan Participation Group, The Johns Hopkins University, the Korean Scientist Group, Los Alamos National Laboratory, the Max-Planck-Institute for Astronomy (MPIA), the Max-Planck-Institute for Astrophysics (MPA), New Mexico State University, University of Pittsburgh, University of Portsmouth, Princeton University, the United States Naval Observatory, and the University of Washington.

\clearpage

\clearpage

\clearpage

\clearpage

\clearpage

\clearpage

\clearpage

\clearpage

\clearpage

\clearpage

\begin{deluxetable}{ccccccccl}
\tabletypesize{\scriptsize}
\tablewidth{0pt}
\tablecaption{Known Quasars in the QVS \label{data}}
\tablehead{
\colhead{Identifier} & \colhead{GCL} & \colhead{$Q_i$} & \colhead{Redshift\tablenotemark{a}} &\colhead{lg(L$_{2500 \AA}$)\tablenotemark{b}} & \colhead{lg(L$_{2 keV}$)} & \colhead{lg(L$_{2 \micron}$)} & \colhead{FIRST F$_{int}$} & \colhead{Cross-identification} \\ (QUEST J...) & & & & & & & \colhead{($mJy$)} & \\}
\startdata
100110.5-004049.0 &  75.42 &  0.82 & 0.13 & 28.52 &  \nodata &  \nodata & \nodata & SDSSJ100110.52-004049.2 \\
100113.6-001234.2 &  77.95 &  1.26 & 2.46 & 31.72 &  \nodata &  \nodata & \nodata & SDSSJ100113.63-001234.5 \\
100143.3-003610.2 &  58.22 &  2.34 & 1.48 & 31.12 &  \nodata &  \nodata & \nodata & SDSSJ100143.28-003610.5 \\
100145.1-011220.6 &  66.71 &  2.20 & 0.46 & 29.72 &  \nodata &  \nodata & \nodata & 2QZJ100145.0-011221 \\
100215.9-001055.7 &  71.77 &  2.00 & 0.35 & 29.55 &  \nodata &  \nodata & \nodata & SDSSJ100215.83-001056.1 \\
100250.0-002453.1 &  59.40 &  1.47 & 0.80 & 31.08 &  \nodata & 31.60 & \nodata & SDSSJ100249.94-002453.5 \\
100253.2-001726.6 &  11.32 &  0.15 & 0.74 & 30.36 &  \nodata &  \nodata & \nodata & 2QZJ100253.2-001728 \\
100255.1-002449.4 & 100.00 &  4.16 & 0.12 & 28.85 & 25.93 & 29.55 & \nodata & SDSSJ100255.11-002449.8 \\
100350.2-005658.2 &  74.92 &  2.92 & 0.79 & 30.44 &  \nodata &  \nodata & \nodata & 2QZJ100350.1-005658 \\
100356.2-005940.6 &  44.70 &  1.05 & 2.11 & 31.84 &  \nodata &  \nodata & \nodata & SDSSJ100356.15-005940.5 \\
\enddata
\tablenotetext{a}{Reported redshift values are from source indicated by Cross-identification.}
\tablenotetext{b}{All luminosity densities are in units of $erg\ s^{-1}\ Hz^{-1}$.}
\tablecomments{Table\ \ref{data} is presented in its entirety in the electronic edition of the Astronomical Journal. A portion is shown here for guidance regarding its form and content.}
\end{deluxetable}

\clearpage

\begin{deluxetable}{rcccc}
\tabletypesize{\small}
\tablewidth{0pt}
\tablecaption{Multiwavelength Luminosities for Quasar Subsets \label{props}}
\tablehead{
\colhead{Sample} & \colhead{N} & \colhead{1st Quartile} & \colhead{median} & \colhead{3rd Quartile}
}
\startdata
& & \multicolumn{3}{c}{L$_{2500 \AA} (10^{31}\ erg\ s^{-1}\ Hz^{-1})$} \\
\hline
Full Sample var & 361 & 0.214 & 0.811 & 2.31 \\
Full Sample all & 933 & 0.363 & 1.19 & 3.26 \\
Full Sample non & 361 & 0.637 & 1.63 & 3.84 \\
\hline
& & \multicolumn{3}{c}{FIRST F$_{int}$ ($mJy$)}\\
\hline
FIRST var & 32 & 3.54 & 9.64 & 42.7 \\
FIRST all & 80 & 2.38 & 6.05 & 30.9 \\
FIRST non & 32 & 1.77 & 3.96 & 27.7 \\
\hline
& & \multicolumn{3}{c}{L$_{2 \micron} (10^{31}\ erg\ s^{-1}\ Hz^{-1})$}\\
\hline
2MASS var & 111 & 0.351 & 1.43 & 6.93 \\
2MASS all & 205 & 0.488 & 2.83 & 21.3 \\
2MASS non & 39 & 1.37 & 9.51 & 26.6 \\
\hline
& & \multicolumn{3}{c}{L$_{2 keV} (10^{26}\ erg\ s^{-1}\ Hz^{-1})$} \\
\hline
ROSAT var & 28 & 0.416 & 1.69 & 7.41 \\
ROSAT all & 38 & 0.540 & 1.98 & 9.38 \\
ROSAT non & 3 & (1.13)\tablenotemark{a} & 3.92 & (12.9)\tablenotemark{a} \\
\enddata
\tablenotetext{a}{ROSAT non-variable population contains 3 quasars. Parenthetical first and third quartile amounts reflect entire range of flux values.}
\end{deluxetable}

\clearpage

\begin{deluxetable}{cccc}
\tabletypesize{\small}
\tablewidth{0pt}
\tablecaption{Quasar Variability Statistics \label{stats}}
\tablehead{ 
\colhead{Sample (N)} & \colhead{GCL $> 93$} & \colhead{$Q_i > 2$} & \colhead{Both} \\ & Only & Only & Tests}
\startdata
QVS (198,213) & 1,610 & 4,130 & 624 \\
Known Quasars (933) & 361 & 604 & 351 \\
\\
\hline
\\
Recovery of Known Quasars & $38.7\%$ & $64.7\%$ & $37.6\%$ \\
Efficiency & $\geq 22.4\%$ & $\geq 14.6\%$ & $\geq 56.3\%$ \\  
\enddata
\end{deluxetable}

\end{document}